\documentclass[twocolumn,showpacs,superscriptaddress,reprint,floatfix,aps,prl]{revtex4-1}

\usepackage{amsmath,amssymb,amsthm}
\usepackage{graphicx}

\usepackage{hyperref}
\hypersetup{
    colorlinks=true,
    linkcolor=cyan,
    filecolor=magenta,      
    urlcolor=blue,
    citecolor=blue,
}

\usepackage{cleveref}

\makeatletter
\newcommand*{\declarecommand}{%
  \@star@or@long\declare@command
}
\newcommand*{\declare@command}[1]{%
  \provide@command{#1}{}%
  \renew@command{#1}%
}
\begin{document}
\makeatother\textbf{}

% bolded letters
\declarecommand{\u}{{\mathbf{u}}}
\declarecommand{\U}{{\mathbf{U}}}
\declarecommand{\f}{{\mathbf{f}}}
\declarecommand{\x}{{\mathbf{x}}}
\declarecommand{\n}{{\mathbf{n}}}
\declarecommand{\X}{{\mathbf{X}}}
\declarecommand{\b}{{\mathbf{b}}}
\declarecommand{\a}{{\mathbf{a}}}
\declarecommand{\p}{{\mathbf{p}}}
\declarecommand{\e}{{\mathbf{a}}}
\declarecommand{\d}{{\boldsymbol{\tau}}}
\declarecommand{\s}{{\alpha}}
% \declarecommand{\hatt}{\mathbf{\hat{t}}}%
\declarecommand{\r}{{\mathbf{r}}}
\declarecommand{\hatt}{{\r_s}}
% bolded symbols
\declarecommand{\bmu}{{\boldsymbol{\mu}}}
\declarecommand{\balpha}{{\boldsymbol{\alpha}}}
% gradient operator
\declarecommand{\grad}{\nabla}
% Identity tensor
\declarecommand{\Id}{\mathbb{I}}
% force  operator
\declarecommand{\F}{{\boldsymbol{\mathcal{F}}}}
% anisotropy operator
\declarecommand{\A}{{\mathbb{A}}}
% averaged quantities
\declarecommand{\au}{{\langle\u\rangle}}
\declarecommand{\ap}{{\langle p\rangle}}
\declarecommand{\auj}{{\langle\u^{-j}\rangle}}
\declarecommand{\apj}{{\langle p^{-j}\rangle}}
\declarecommand{\af}{{\langle\f\rangle}}

\declarecommand{\blue}[1]{{\color{blue}#1}}

%\begin{document}
\title{Swirling Instability of the Microtubule Cytoskeleton}
\date{\today}
\author{David B. Stein}
\thanks{These authors contributed equally}
\affiliation{Center for Computational Biology, Flatiron Institute, 162 5th Ave., New York, NY 10010}
\author{Gabriele De Canio}
\thanks{These authors contributed equally}
\affiliation{Department of Applied Mathematics and Theoretical Physics, Centre for Mathematical 
Sciences, University of Cambridge, Cambridge CB3 0WA, United Kingdom}
\author{Eric Lauga}
\email{e.lauga@damtp.cam.ac.uk}
\affiliation{Department of Applied Mathematics and Theoretical Physics, Centre for Mathematical 
Sciences, University of Cambridge, Cambridge CB3 0WA, United Kingdom}
\author{Michael J. Shelley}
\email{mshelley@flatironinstitute.org}
\affiliation{Center for Computational Biology, Flatiron Institute, 162 5th Ave., New York, NY 10010}
\affiliation{Courant Institute, New York University, 251 Mercer St., New York, NY 10012}
\author{Raymond E. Goldstein}
\email{R.E.Goldstein@damtp.cam.ac.uk}
\affiliation{Department of Applied Mathematics and Theoretical Physics, Centre for Mathematical 
Sciences, University of Cambridge, Cambridge CB3 0WA, United Kingdom}

\title{Swirling Instability of the Microtubule Cytoskeleton}
\date{\today}

\begin{abstract}
In the cellular phenomena of cytoplasmic streaming, molecular motors carrying cargo 
along a network of microtubules entrain the surrounding fluid.
The piconewton forces produced by individual motors are sufficient to deform 
long microtubules, as are the collective fluid flows generated by many moving motors.  
Studies of streaming during oocyte development in the fruit
fly \textit{D. melanogaster} have shown a transition from a spatially-disordered 
cytoskeleton, supporting flows with only 
short-ranged correlations, to an ordered state with a cell-spanning vortical flow.
To test the hypothesis that this transition is driven by fluid-structure interactions 
we study a discrete-filament model and a coarse-grained continuum theory 
for motors moving on a deformable cytoskeleton, both of which 
are shown to exhibit a \textit{swirling instability} to spontaneous large-scale rotational 
motion, as observed.

\end{abstract}
\maketitle

\begin{figure}[b]
    \includegraphics[clip=true,width=0.98\columnwidth]{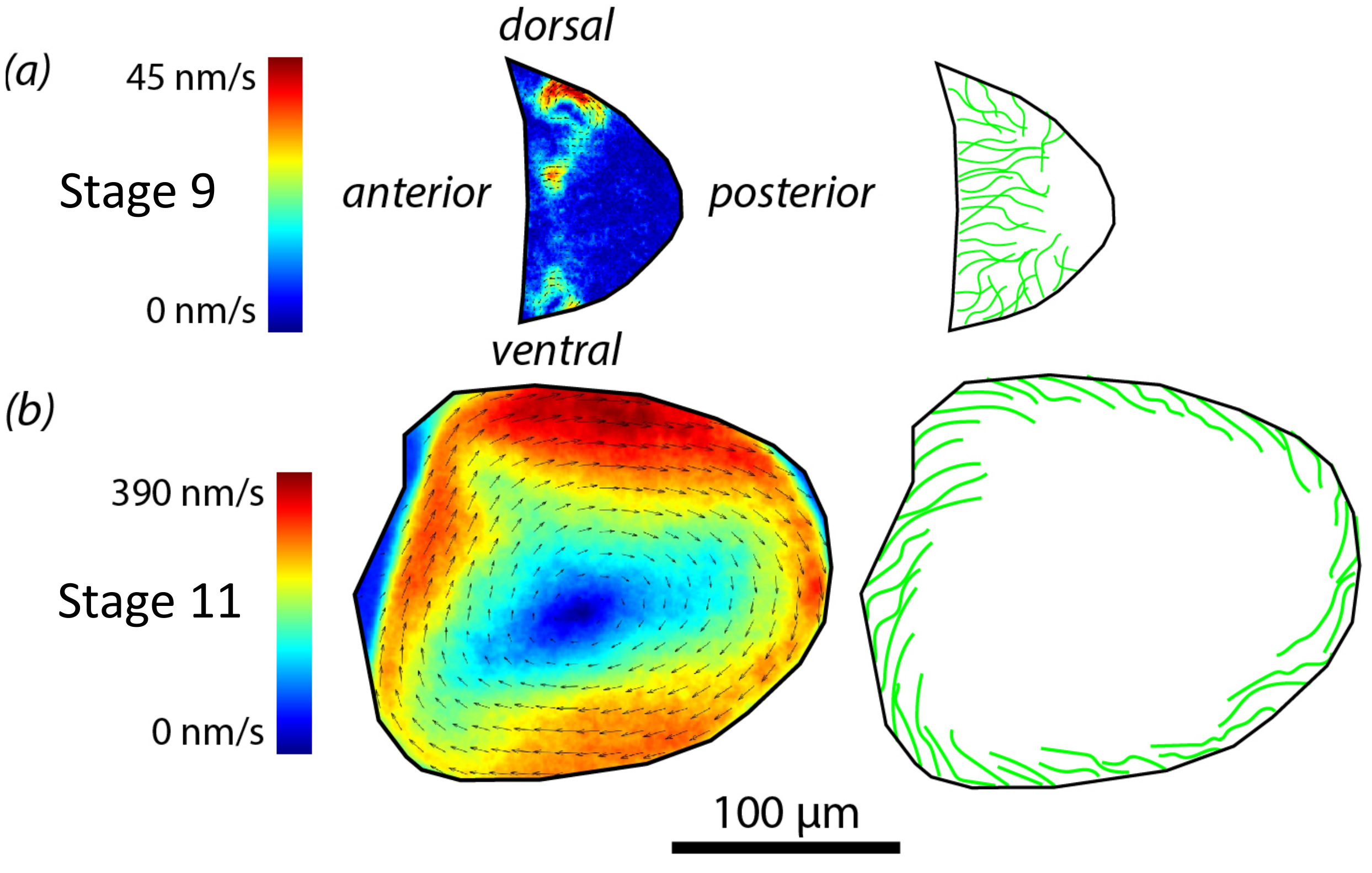}
    \caption{Cytoplasmic streaming flows in the \textit{Drosophila} oocyte.  (a) Experimental 
    flow field \cite{flowfield}
    and schematic of the disordered swirling flows and microtubule organization in early stages
    of development.  (b) Later flows organize into a single vortex as MTs lie parallel to the cell 
    periphery.}
    \label{fig1}
\end{figure}

A striking example of fluid-structure interactions within cells 
\cite{NeedlemanShelley} occurs in oocytes of the fruit fly 
\textit{Drosophila melanogaster} \cite{Drosophila_review1}.  These develop over 
a week from a single cell through repeated rounds of cell division, 
differentiation and growth, ultimately reaching hundreds of microns across.  
This pathway has historically
been divided into $14$ stages, and it is in stages $9-11$, at days $6.5-7$ \cite{HeWangMontell}, 
that fluid motion is most noticeable.  In stage $9$ (Fig.~\ref{fig1}), microtubules (MTs) reach
inward from the oocyte periphery, forming a dense assembly along which 
molecular motors (kinesins) move at tens of nm/sec, carrying messenger 
RNAs and other nanometric particles.  This motion 
entrains the surrounding fluid, producing cytoplasmic streaming \cite{streaming,vandeMeent_review}
that can be visualized several ways: in brightfield by the motion of endogenous 
particles \cite{Gutzeit,Theurkauf92,Theurkauf94}, via their
autofluorescence \cite{Palacios02,Serbus05}, and through a combination of particle image velocimetry
and fluorescently labelled microtubules 
\cite{Ganguly,Drechsler1,Drechsler2}.
Previous work \cite{Theurkauf92,Ganguly}
revealed that these flows initially take the form of transient, recurring
vortices and jets whose correlation length is a fraction of the cell scale, with 
no long-range order.  But by stage $11$, a dramatic reconfiguration of the cytoskeleton occurs, coincident with the appearance of a vortex 
spanning the entire cell \cite{Gutzeit,Theurkauf92,Serbus05,Monteith}. 

Kinesin motors move from the {\it minus} ends of microtubules (attached to the 
oocyte periphery) to the {\it plus} ends (free in the interior).
Transport of cargo through the network 
depends on motor-microtubule binding details \cite{KhucTrong_PRL,Williams} and 
on the mesh architecture \cite{KhucTrong_eLife,Gelfand_jcb}.  As a motor pulls cargo toward the plus 
end the filament experiences a localized minus-end-directed compressive force,
as in Euler buckling.  For a filament of length $L$ and 
bending modulus $A$ \cite{Gittes}, 
the buckling force is $\sim \pi A/L^2\sim 60\,{\rm pN}/L^2$, 
where $L$ is measured in microns.  Thus, a 
kinesin's force of several pN \cite{motorforce} can buckle 
MTs $10-40\,\mu$m long.  

The coupled filament-motor problem is richer than
Euler buckling because a motor exerts a ``follower force" \cite{followerforce}  
that is aligned with the filament.  This feature breaks the 
variational structure of the problem and allows a filament pinned at its minus end to 
oscillate even at zero Reynolds number \cite{BaylyDutcher,deCanio,kanso_follower}.  
By exerting a force on the fluid a motor induces long-range flows 
which, if compressive, can further deform 
filaments \cite{YoungShelley,Kantsler}.

\begin{figure*}[t]
    \includegraphics[clip=true,width=1.95\columnwidth]{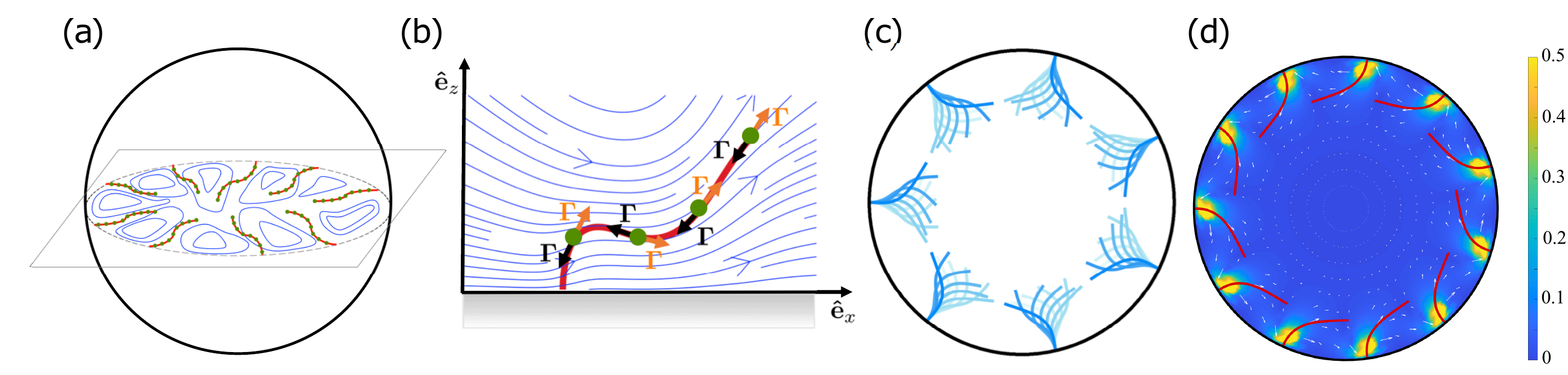}
    \caption{Discrete filament computations.  (a) $N$ equally spaced elastic filaments clamped at their
    attachment points, reach inward from a no-slip spherical shell. Each has a continuous distribution of 
    tangential point forces (red)
    that (b) exert a force ${\bf \Gamma}$ on the fluid and an equal and opposite 
    compressive force on the filament. Synchronous oscillations ($N=7$, 
    $\sigma=1700$), (d) steady, bent configuration ($N=9$, $\sigma=500$) and swirling flow field.}
    \label{fig2}
\end{figure*}

It has been hypothesized \cite{Serbus05,Monteith} that the transition from disordered 
streaming flows to a single vortex in stage $11$ is a consequence of the kinds of fluid-structure 
interactions described above, facilitated by a decrease in cytoplasmic viscosity that accompanies
the disappearance of a coexisting network of the biopolymer f-actin.
Here, through a combination of direct computations on the coupled filament-flow problem 
\cite{deCanio}
and studies of a recent continuum theory for dense filament suspensions 
\cite{SteinShelley}, we confirm this hypothesis 
by showing the existence of a novel {\it swirling instability} of the cytoskeleton.

The swirling instability can be understood in a 
simplified model of the oocyte: a rigid sphere of radius $R$ containing a fluid of viscosity
$\mu$, with $N$ elastic filaments reaching inwards from clamped attachment points
equally spaced around the equator.  A slice in the filament plane 
(Fig.~\ref{fig2}(a)) appears like a confocal slice of the oocyte (Fig.~\ref{fig1}).  
The filaments have a radius $r$, a constant length $L$, bending modulus $A$ and a uniform line density $f$ of 
follower forces (Fig.~\ref{fig2}(b)).  Although free microtubules have a complex 
dynamics of growth and decay, recent evidence \cite{Gelfand} for 
`superstable' cortically-bound microtubules in stages displaying unidirectional 
streaming justifies the constant-length approximation.

Microtubules are the quintessential slender bodies \cite{KellerRubinow} of biophysics, with aspect 
ratios $\varepsilon=r/L$ of ${\cal O}(10^{-3})$.  As their self-interactions are weak, we use local slender-body theory \cite{RFT,TornbergShelley} to obtain the dynamics.
In an arclength parameterization $s$, 
the $j^{\rm th}$ filament ${\bf r}^j(s,t)$ evolves as
\begin{equation}
    \eta\left(\r_t^j - \U^j\right) = (\mathbb{I} + \r_s^j\r_s^j)\left(-A\r_{4s}^j + (\Lambda^j\r_s^j)_s + f\r^j_s\right),
    \label{eom1}
\end{equation}
where $\r^j_s$ is the unit tangent, $\eta=8\pi\mu/c$, with $c=|\ln (e\varepsilon^2)|$, and the
Lagrange multiplier 
$\Lambda^j$ enforcing inextensibility obeys a second-order PDE \cite{GL}. 
In the background flow ${\bf U}^j={\bf u}^j+{\bf u}^{i\to j} + 
{\bf v}^{i\to j}$, ${\bf u}^j$ is that produced by the motors on
$j$, ${\bf u}^{i\to j}$ is due to the motors on
$i\ne j$, and ${\bf v}^{i\to j}$ is due to motion of filaments $i\ne j$.
For example, the induced flow due to the $j$th fiber is ${\bf u}(\x)=\int_0^L ds 
f \r_s^j(s)\cdot {\bf G}(\x-\r^j(s))$ (see Supplemental Material \cite{SM,thesis}), with ${\bf G}$ the Greens 
function appropriate to the interior of a no-slip sphere \cite{MaulKim}.
Filament clamping at the sphere implies that $\r^j(0,t)$ remains fixed and that
$\r_s^j(0,t)$ is the inward sphere normal at the attachment point. The free end is torque- and force-free: $\r^j_{ss}(L,t)=\r^j_{sss}(L,t)=\Lambda(L,t)=0$.

A single fiber clamped at a flat wall displays
a supercritical 
Hopf bifurcation which, expressed in terms of the dimensionless motor force $\sigma\equiv fL^3/A$, occurs 
at $\sigma^*\simeq 124.2$, beyond which the
filament exhibits steady oscillations 
whose amplitude grows as $\sqrt{\sigma-\sigma^*}$ \cite{deCanio}.  
When several filaments interact within the sphere (\ref{fig2}c) they 
also oscillate, but with their motions synchronized
in phase, very much like eukaryotic 
flagella \cite{synchro}.  The dynamical model \eqref{eom1} contains
two ingredients often found necessary for such synchronization \cite{NEL}: hydrodynamic interactions and
the ability of a filament to change shape and thereby 
adjust its phase in response to those flows.

As the filament density and motor strength are increased we find the swirling instability: a
transition to a {\it steady} configuration of bent filaments whose distal parts are oriented almost parallel
to the wall (Fig.~\ref{fig2}(d)).  The bent configuration is maintained 
by azimuthal flows,
induced by the motors, that generate drag along the 
distal part of the filaments, and thus a torque opposing the bending torques near the filament base.
As with any such spontaneous breaking of symmetry,
both left- and right-hand configurations are possible; the choice between the two is dictated by
initial conditions.  This transition is reminiscent of the self-organized
rotation of cytoplasmic droplets extracted from plants \cite{Yotsuyanagi} and 
the spiral vortex state of confined bacterial suspensions \cite{confinement_expt}, both
modeled as suspensions of stresslets \cite{Saintillan,Saintillan2018extensile,confinement_theory}.

\begin{figure}[t]
    \includegraphics[clip=true,width=\columnwidth]{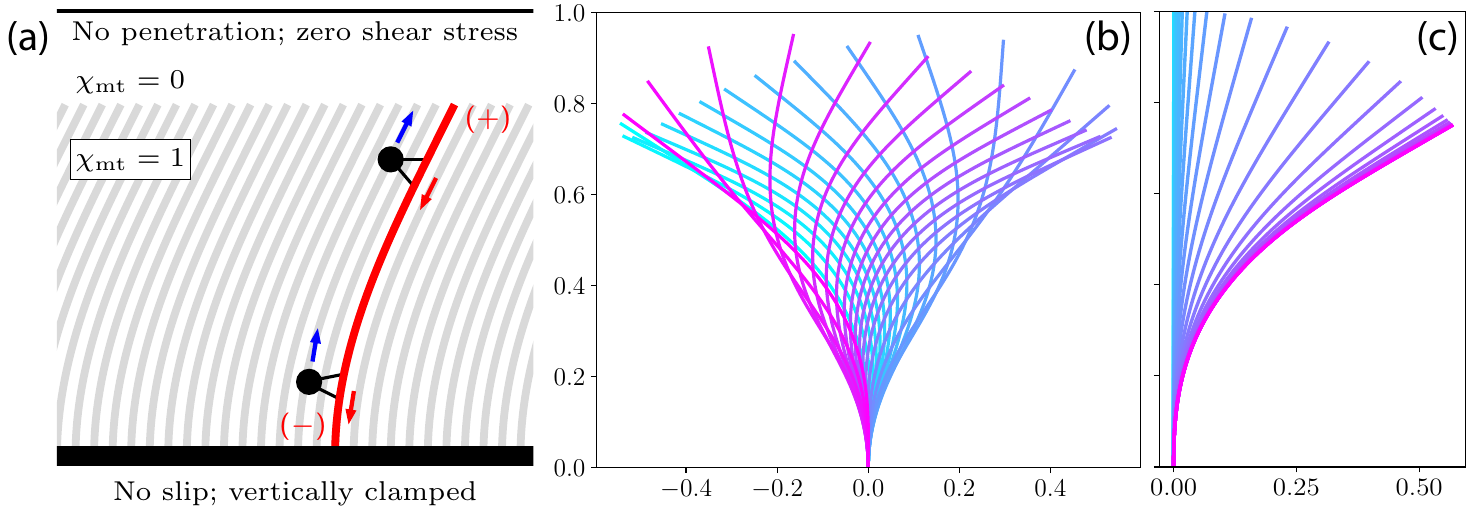}
    \caption{Continuum model in planar geometry. (a) Bi-infinite array of 
    MTs, whose minus ends are clamped vertically at a no-slip boundary. Results of full 
    computations at (b) $\rho=4.65$, $\sigma=70$ and (c) $\sigma=39$. Colors  
    denote time, from cyan (early) to pink (late).}
    \label{fig3}
\end{figure}

While direct computations on denser arrays of discrete filaments are possible \cite{NRZS}, cortically 
bound oocyte microtubules are so tightly packed, with an inter-fiber spacing 
$\delta\ll L$ \cite{Serbus05,Ganguly,Drechsler1,Drechsler2}, that a continuum approach is
justified.  The description we use \cite{SteinShelley}, in which microtubules form an 
anisotropic porous medium, is based on the map $\X=\r(\balpha)$,
where the Lagrangian coordinate $\balpha=(\alpha,s)$ encodes the location $\alpha$ 
of the minus ends of the microtubules and arclength $s$.  
In a system of units made
dimensionless by the length $L$ and elastic relaxation time $\eta L^4/A$,
we obtain a continuum version of \eqref{eom1},
\begin{equation}
     \r_t - \u|_{\r(\balpha)}  = (\mathbb{I}+\hatt\hatt)\cdot\left(-\r_{ssss} + (\Lambda\hatt)_s 
     - \sigma\hatt\right)\,.
     \label{eq:continuum:fiber}
\end{equation}
The fluid velocity 
$\u$ arises from the force distribution along the filaments and is evaluated at the
Eulerian position $\x$ according to an inhomogeneous Stokes equation,
\begin{equation}
     -\nabla^2\u + \grad p = \chi_\textnormal{mt}\rho\left[\mathcal{J}^{-1}(-\r_{ssss} + (\Lambda\r_s)_s)\right]|_{\r^{-1}(\x)}\,,
     \label{eq:continuum}
\end{equation}
subject to the incompressibility constraint $\grad\cdot\u = 0$. The indicator function $\chi_\textnormal{mt}$ is 
supported where the MT array is present (Fig.~\ref{fig3}a).  Here, $\rho=8\pi\rho_0 L^2/c$ is the
rescaled areal number density of microtubules, expressible as $\rho=\phi(L/\delta)^2$, where the constant
$\phi$ depends only on the MT slenderness and packing geometry at the wall; 
$\phi\approx4$ when $c\approx10$ and the MTs are hexagonally packed.
The quantity $\mathcal{J}=\det[\partial\r/\partial\balpha]$ measures the change in 
microtubule density due to deformations of the array; $\mathcal{J}^{-1}$ increases as fibers move closer together.

The simplest geometry is an infinite planar array of MTs, as shown in  
 Fig.~\ref{fig3}(a).  As in the discrete model, the MTs are normally 
clamped to a no-slip wall and are 
force- and torque-free at their plus ends. At a distance $H$ above the wall, 
no-penetration and zero-tangential stress conditions are imposed on the fluid. For dynamics homogeneous 
along $x$, the fluid flow is unidirectional and constant above the MTs, so $H$ plays no role.
Nonlinear computations \cite{methods} reveal both oscillatory dynamics and the emergence of steady streaming. 
Fig.~\ref{fig3}(b) shows the dynamics when $\rho=4.65$ and $\sigma=70$: self-sustaining oscillations of the 
MT array are observed, similar to those in Fig.~\ref{fig2}(c). Note that while Fig.~\ref{fig3}(b) shows only a single filament, it represents the common dynamics of 
{\it all} of the collectively beating filaments in the array. When $\sigma$ is decreased to $\approx39$, the 
MT array deforms and stabilizes into a \textit{steady} 
bent state (Fig.~\ref{fig3}(c)). This 
represents the continuum description of the swirling 
transition, with similar dynamics to those observed in the discrete results.  

An equilibrium of the system occurs when filaments are aligned straight along $z$, with $\u=0$ and $\Lambda=-\sigma(1-z)$. For $\sigma>0$, the motor-force is 
compressive and buckling may occur.  
A small transverse perturbation in fiber shape of the form  $\hatt={\bf \hat z}+\epsilon g(z){\bf \hat x}$ ($\epsilon\ll1$) evolves as 
\begin{equation}
   g_t = -g_{zzzz} -\sigma\left[(1-z)g_{z}\right]_z + \rho
   \left[\sigma(1-z)g+g_{zz}\right].
    \label{linstab}
\end{equation}
The first two terms are like those of an elastic filament under an aligned gravitational load, with an internal tension varying linearly
from one end to the other \cite{Landau,ponytails}. The third is the fiber forces filtered through the 
non-local Stokes operator, capturing hydrodynamic interactions within the fiber array (and hence the $\rho$ prefactor). That this term is local is both fortunate and surprising, and follows from the simplicity of the 
Stokes flow in this case. The term $\rho g_{zz}$ captures the additional resistance to bending from flow: if a MT is to bend, it must move the 
fluid around it, bending other MTs; the term $\rho\sigma(1-z)g$ is destabilizing: if a 
MT is to remain straight, it must resist the fluid motions generated by MTs around it.

\begin{figure*}[t]
    \includegraphics[clip=true,width=2.0\columnwidth]{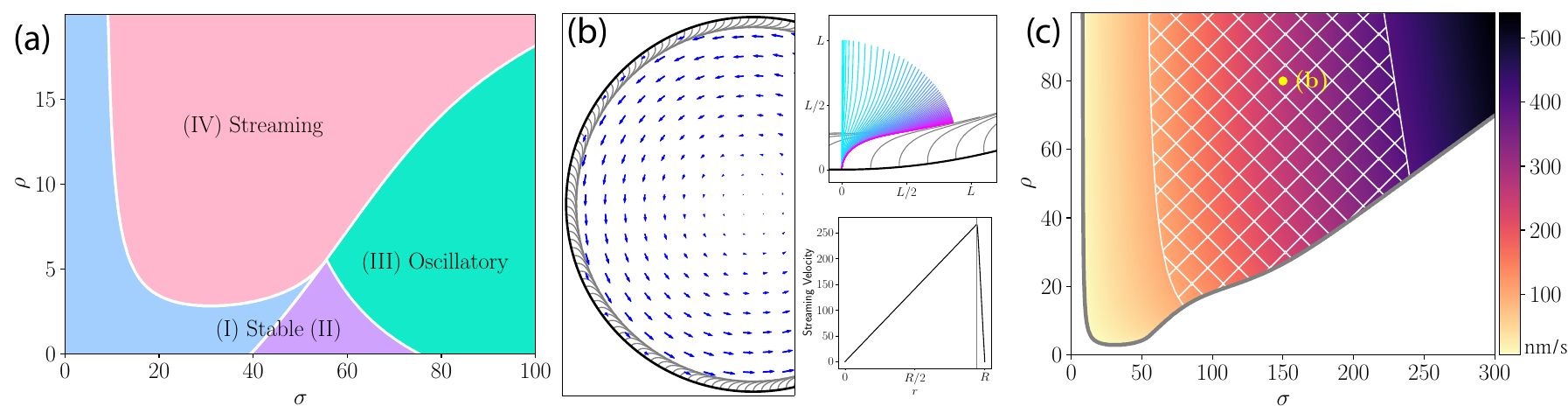}
    \caption{Continuum model in cylindrical geometry. (a) Results of linear stability analysis about the radially aligned 
    state, with $R=5L$. 
    (b) Steady-state fiber deformations and velocity field for $\sigma=150$ and $\rho=80$. 
    Density of visualized fibers corresponds to the physical density. Top inset shows deformed MTs and the dynamics of 
    a representative one (see also supplemental video \cite{SM}). Bottom inset shows the azimuthal velocity field as a function of $r$. 
    (c) Dimensional streaming velocities in parameter space; hatched region is consistent
    with \emph{in vivo} estimates of $100-400\,$nm/s. Yellow dot denotes simulation shown in (b).}
    \label{fig4}
\end{figure*}

While the planar geometry reproduces all qualitative features of the streaming transition \cite{SM}, to capture
the key feature of confined hydrodynamic interactions in the oocyte we extend the analysis to a cylindrical domain, 
where the no-flow steady state is an array of straight MTs 
pointing inwards. Fig.~\ref{fig4} shows the results of a linear stability analysis for an experimentally 
relevant ratio of cylinder diameter to MT length of $10:1$. 

For $\rho\ll1$, the continuum model behaves like isolated fibers with negligible collective fluid entrainment. 
For small $\sigma$, straight fiber arrays are stable (regions I $\&$ II, with region II having oscillatory decay to 
equilibrium), but with increasing $\sigma$ there is a Hopf bifurcation to a state that nonlinear simulations 
show has oscillations (cf. Fig.~\ref{fig2}(c)). For $\rho\gtrsim2.8$ ($\delta\lesssim 1.2L$), a new region of 
instability (IV) appears, with real and positive eigenvalues; nonlinear simulations show this 
leads to collective MT bending and swirling flows.  

Figure \ref{fig4}(b) shows a nonlinear simulation of the transition to streaming
in region IV.  The upper inset shows the development of the instability,
with successive MTs bending over to form a dense canopy above their
highly curved bases. At steady state, the concentrated motor forces within
the canopy are azimuthally aligned, almost a $\delta$-function a distance $\sim L/4$ above the
wall, and drive the large-scale
streaming flow. The ooplasmic flow beneath the MT canopy is nearly a linear shear flow, 
transitioning above to solid body rotation, the solution to Stokes flow forced at a
cylindrical boundary.

We now estimate ranges of density and force that are consistent with 
observed streaming speeds $u\approx 100-400\,$nm/s (Fig.~\ref{fig1} and \cite{Gelfand,Monteith}). Taking $L=20\,\mu$m, 
$\mu=1\,$ Pa s \cite{Ganguly} and $A=20\,$pN$\mu$m$^2$, we obtain a velocity scale $A/\eta L^3 \approx 1\,$nm/s and a force-density scale $A/L^3\approx 2.5\,$fN/$\mu$m. 
Figure \ref{fig4}c shows the streaming speeds calculated from nonlinear 
simulations in region IV. Those with maximum speeds falling in the experimental range  
lie in the hatched area. Increasing $\rho$ only marginally increases streaming speeds, and so to increase flow speed while remaining in region IV requires increasing both $\rho$ and $\sigma$. The minimum value of $\rho\approx20$ that is consistent with observed streaming velocities corresponds to $\delta\lesssim 0.4 L$, a more stringent constraint than that required for the streaming transition. The force densities consistent with streaming speeds are $f\!\sim\! 0.1-0.6\,$pN/$\mu$m. Speeds on the higher end of the physical range approach the $\approx\! 700\,$nm$/$s of kinesin-1 under negligible load \cite{motorforce}, while cargo speeds on oocyte MTs are $200-500\,$nm$/$s \cite{Monteith,Gelfand,Loiseau}. Assuming a linear force-velocity relation and a stall force of $6\,$pN \cite{motorforce} gives a single motor force of $\approx\! 2\,$pN; approximately $1-6$ kinesins are required per $20\, \mu$m MT to generate these force densities.

It may be surprising that the streaming speed only weakly depends on $\rho$. A heuristic argument for the flow speeds views the cytoskeleton as a porous medium of permeability 
$k\sim \delta^2$, in which speed $u\sim (k/\mu)\nabla p$, where the pressure gradient (force/volume) from motors is $f/\delta^2$, yielding $u\sim f/\mu \sim (A/\eta L^3)(8\pi/c)\sigma$, independent of $\rho$.  This relationship is surprisingly accurate \cite{SM}.

\begin{figure}[b]
    \includegraphics[clip=true,width=0.95\columnwidth]{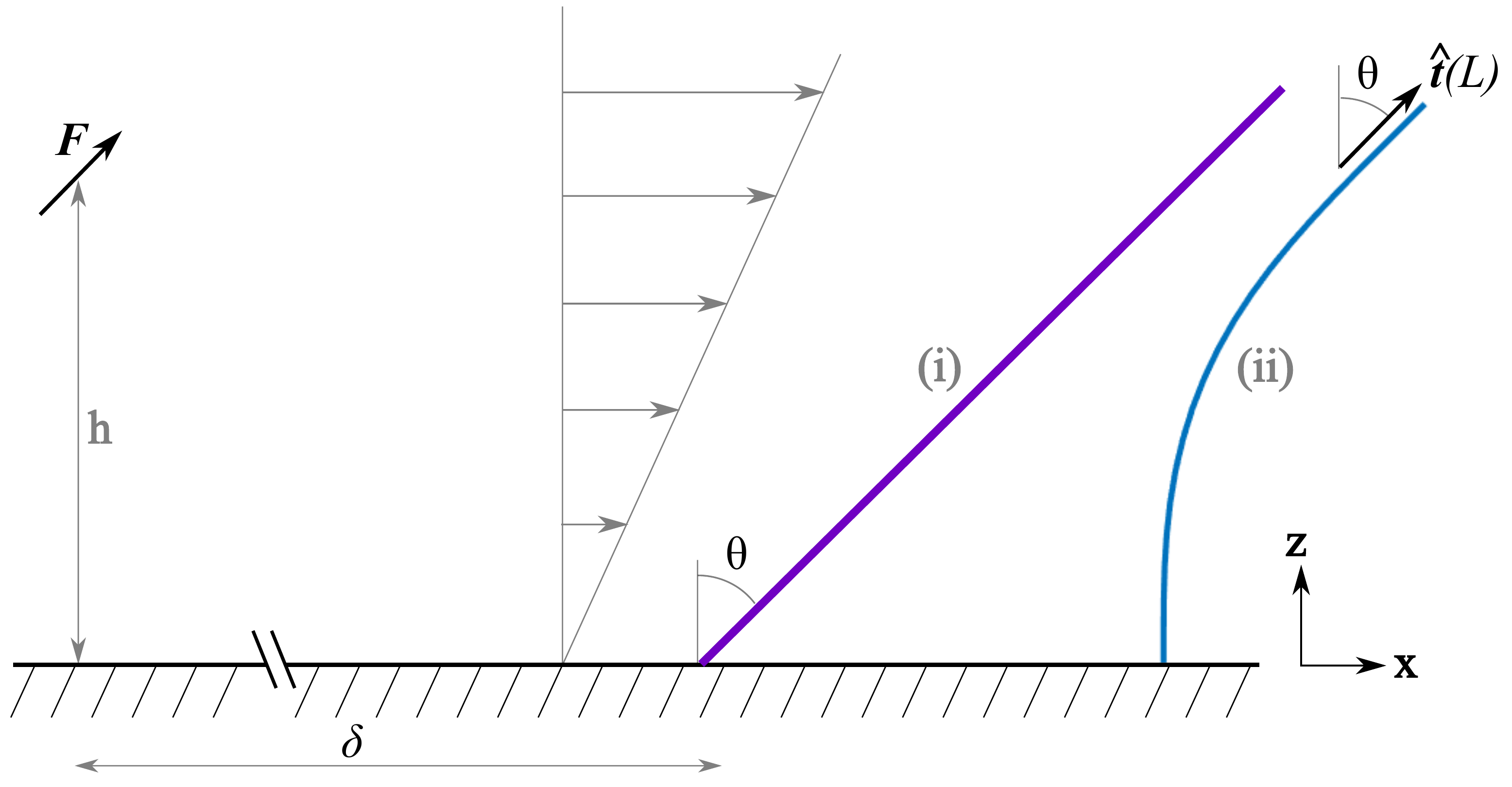}
    \caption{Self-consistent model.  An upstream point force ${\bf F}$ 
    parallel to the distal end of a filament produces shear flow that 
    deflects the filament. Two variants of the model: (i) rigid rod
    with a torsional spring at its base, (ii) a clamped elastic filament.}
    \label{fig5}
\end{figure}

When the density $\rho$ is sufficiently high, the swirling instability first appears for force densities $\sigma$ 
substantially smaller than those that induce buckling instabilities in a single filament. Thus this transition 
must be driven by the additional hydrodynamic destabilization that neighboring fibers impart (in the simplest geometry, 
given by the term $\rho\sigma(1-z)g$ in Eq.~\ref{linstab}). This observation motivates a simple heuristic argument 
for the instability, in which a filament is bent by the flow produced primarily 
by its upstream neighbor, whose distal half is nearly parallel to the wall.  
Seen from a distance, that bent portion acts on the fluid like a point force \cite{eLife} 
${\bf F}\sim (fL/2)\r_s(L)$ oriented along its distal tangent vector 
(Fig.~\ref{fig5}), displaced a distance $h\sim L/2$ 
from the surface.  Near a no-slip wall, the far-field flow 
along $x$ due to a force ${\bf F}\!\parallel\!{\bf \hat x}$ a distance $\delta$ upstream is simple shear 
\cite{Blake,lauga_pillars},
\begin{equation}
    {\bf U}(x,z)=\dot\gamma z {\bf \hat e}_x\,,
    \label{shearflow}
\end{equation}
where $\dot\gamma=3h F/2\pi\mu \delta^3$. Self-consistency requires the magnitude of the force driving the shear be 
given by the projection of ${\bf F}$ along $x$, so $\dot\gamma \to \dot\gamma\sin(\theta(L))$.

The very simplest model to illustrate the self-consistency condition 
is a rigid MT with a torsional spring at its base  that provides a restoring
torque $-k\theta$ (Fig.~\ref{fig5}(i)).  
With $z(s)=s\cos\theta$ and $\eta {\bf \hat n}{\bf \hat n}\cdot {\bf U}$ 
the local normal force on a segment, the local torque about the point $s=0$ is 
$\eta\dot\gamma s^2 \cos^2\!\theta$ which, when integrated along the filament and balanced
against the spring torque, yields the 
self-consistency condition 
\begin{equation}
    \theta=B\sin\theta\cos^2\!\theta\,,
\end{equation}
where $B=\eta \dot\gamma L^3/k$.
For $B<1$ (slow flow or a stiff spring) $\theta=0$ is the only fixed point, while for 
$B\gtrsim 1$ two mirror-image swirling solutions appear through a pitchfork bifurcation, 
$\theta_{\pm}\simeq (6(B-1)/7)^{1/2}$.

To study the interplay between filament oscillations and swirling we use \eqref{shearflow}
in the filament dynamics \eqref{eom1}, where the control parameter for the shear flow is \cite{YoungShelley,Kantsler}
\begin{equation}
    M=\frac{\eta\dot\gamma L^4}{A}\sim \frac{3\sigma}{c}\left(\frac{\rho}{\phi}\right)^{3/2},
    \label{Mparameter}
\end{equation}
where the second relation uses the above estimates for $F$ and $h$.  
Since a clamped elastic filament
behaves like a torsional spring with a spring constant $k=A/L$, we see consistency with the
parameter $B$ defined above. 
A numerical self-consistent calculation confirms the existence of a swirling instability \cite{SM}.

Through simplified discrete and continuum models we have demonstrated the existence of
a novel swirling instability of arrays of elastic filaments, thus lending support to the hypothesis 
\cite{Monteith}
that cytoplasmic streaming flows in \textit{Drosophila} 
oocytes are tied to self-organization of the microtubule cytoskeleton.  Future studies could shed light on the detailed mechanism involved in the untangling of the 
\textit{Drosophila} oocyte cytoskeleton when it transitions to the vortical state,
and the possibility of reproducing this transition \textit{in vitro}. Lastly, this study highlights the role of active force dipoles in the self-organization of fluid-biopolymer systems \cite{Saintillan,Saintillan2018extensile,confinement_theory}.

\begin{acknowledgments}
We are indebted to Maik Drechsler and Isabel Palacios for sharing the data in Fig.~\ref{fig1} and to them, Daniel St Johnston and Vladimir Gelfand for discussions on \textit{Drosophila} streaming. 
This work was supported in part by ERC Consolidator grant 682754 (EL), 
Wellcome Trust Investigator Award 207510/Z/17/Z, 
Established Career Fellowship EP/M017982/1 from the
Engineering and Physical Sciences Research Council, and the Schlumberger Chair Fund (REG). MJS acknowledges the support of NSF Grant DMS-1620331.

\end{acknowledgments}

\vfil
\eject

\section{Supplemental Material}

This file contains calculational details for computations of discrete filaments
and the continuum model, further results from linear stability analyses in the
latter, and quantitative analysis of streaming speeds in the swirling state.

\setcounter{equation}{0}
\setcounter{figure}{0}
\setcounter{table}{0}
\setcounter{page}{1}
\makeatletter
\renewcommand{\theequation}{S\arabic{equation}}
\renewcommand{\thefigure}{S\arabic{figure}}
\renewcommand{\bibnumfmt}[1]{[S#1]}
\renewcommand{\citenumfont}[1]{S#1}
%%%%%%%%%% Prefix a "S" to all equations, figures, tables and reset the counter %%%%%%%%%%

\section{Calculational Details for Discrete Filaments}

\subsection{Image system inside a sphere}

By modelling the motor-cargo ensemble with fluid-entraining
follower forces and
approximating the system geometry by a spherical container,
we take advantage of the analytical solution \cite{MK94_SM} for the velocity field
of a point force inside a sphere with a no-slip wall.  
These simplifications allow us to compute the
background flow directly,
instead of numerically solving the
Stokes equations for the fluid flow.
For a point force of magnitude $F$ located at $\x_0$ inside a sphere of unit radius,
the m$^{\text{th}}$ component of the velocity is
\begin{equation}
u_{\text m} = \frac{F_{\text j}}{8 \pi \mu} \left[ G_{\text{jm}}
                            \left(\x - \x_0 \right)  +
                            \bar {G}_{\text{jm}} \left(\x \right) \right] \, ,
\label{vel_sphere}
\end{equation}
where $\mu$ is the dynamic viscosity
of the fluid, $G_{\text{jm}}$ is the Green's function
\begin{equation}
G_{\text{jm}} \left( \x-\x_0 \right)  =
\frac{\delta_{\text{jm}}}{r} + \frac{( x_{\text j} - x_{0_{\text j}} )
                        \left(  x_{\text m} - x_{0_{\text m}} \right)}{r^3} \, ,
\end{equation}
and
\begin{widetext}
\begin{align}
\bar {G}_{\text{jm}} \left(\x\right)   =  & \ \hat e_{\text m} \hat e_{\text k} 
                        \Bigg\{ \frac{1-3R^2_0}{2R^3_0} \  G_{\text{jk}} \left(\x - \bar\x_0 \right)  
                            -\frac{1-R^2_0}{R^4_0} \ \hat e_{\text l} \   G_{\text{jk},l} \left(\x - \x_0 \right) 
                         -\frac{(1-R^2_0)^2}{4R^5_0} \ \nabla^2  G_{\text{jk}} \left(\x - \bar\x_0 \right) \Bigg\} \notag \\
                        & + \left( \delta_{\text{km}} -\hat e_{\text k} \hat e_{\text m} \right)
                        \Bigg\{  \frac{3R^2_0-5}{2R^3_0} \  G_{\text{jk}} \left(\x - \bar\x_0 \right) 
                        +\frac{(1-R^2_0)^2}{4R^5_0} \ \nabla^2  G_{\text{jk}} \left( \x - \bar\x_0 \right) \Bigg\} \notag \\
                        & + \frac{1-R^2_0}{R^4_0} \ \hat e_{\text k}  \left( \delta_{\text{lm}} -\hat e_{\text l} \hat e_{\text m} \right)
                              G_{\text{jk},l} \left(\x - \bar\x_0 \right) 
                         -\frac{3(R^2_0-1)}{R^3_0}  \frac{\left( \delta_{\text{jm}} -\hat e_{\text j} \hat e_{\text m} \right)}{\bar r}
                        +\left(R^2 - 1 \right) \left( \delta_{\text{km}} -\hat e_{\text k} \hat e_{\text m} \right)
                        \frac{\partial \varphi_{\text k}}{\partial x_{\text j}} \, ,
\end{align}
\end{widetext}
with
\begin{equation}
\varphi_{\text k} = - \frac{3(R^2_0 - 1)}{2R^3_0}\ \frac{x_{\text k}}{\bar r} \ \frac{R-\bar R_0 \cos \alpha +
\bar r \cos \alpha}{R \bar R^2_0 \sin^2\alpha} \, .
\end{equation}
Here, $R_0$ and $\bar R_0$ are the norms of $\x_0$ and its image
${\bar \x}_0$, respectively, ${\hat \e} = \x_0/R_0$ is the unit vector in the axial direction,
$r = |\x-\x_0|$, $\bar r = |\x-\bar \x_0|$,
$R$ is the norm of $\x$, and $\alpha$ is the
angle between $\x$ and $\hat \e$.

\subsection{Geometrical formulation}

For both the discrete-filament computations and the self-consistent model, we use the tangent angle representation of the 
filament.  For a filament of length $L$, parameterized by arclength $s$, 
anchored at wall in the 
$x-y$ plane, with $z$ orthogonal to the plane, into the fluid, the position vector is ${\bf r}(s)=(x(s),z(s))$, 
and the unit tangent and normal vectors ${\bf \hat t}(s)={\bf r}_s$ and ${\bf \hat n}(s)$
are
\begin{eqnarray}
    {\bf \hat t}(s)&=&\cos\theta(s){\bf \hat e}_x + \sin\theta(s){\bf \hat e}_z\\
    {\bf \hat n}(s)&=&\sin\theta(s){\bf \hat e}_x -\cos\theta(s){\bf \hat e}_z~,
\end{eqnarray}
where $\theta(s)$ is the angle the tangent vector makes with respect to the $x$-axis.
The curve is obtained by integration of ${\bf \hat t}$, and assuming $(x(0),z(0))=(0,0)$, we have
\begin{equation}
    x(s)=\int_0^s \! ds' \cos\theta(s')~, \ \ \ z(s)=\int_0^s\! ds' \sin\theta(s')~.
\end{equation}
It follows from the above that 
\begin{equation}
    \frac{\partial{\bf \hat t}}{\partial t}=-\theta_t{\bf \hat n}~,
\end{equation}
from which we can obtain the equation of motion for $\theta$ given that for ${\bf r}$.

The equation of motion is that of elastohydrodynamics within resistive force theory,
\begin{equation}
        \left(\zeta_{\parallel} {\bf \hat t}{\bf \hat t}+\zeta_{\perp}{\bf \hat n}{\bf \hat n}\right)
    \cdot\left({\bf r}_t-{\bf U}\right)=-A{\bf r}_{ssss}-\left(\Lambda{\bf r}_s\right)_s~,
    \label{ehd}
\end{equation}
where $A$ is the filament bending modulus, $\Lambda$ is the Lagrange multiplier, and 
${\bf U}=U{\bf \hat e}_x+V{\bf \hat e}_z$ is the background flow experienced by the filament.  
We express the derivatives in \eqref{ehd} in terms of the curvature $\kappa=\theta_s$, using the
Frenet-Serret equations ${\bf \hat t}_s=-\kappa {\bf \hat n}$ and ${\bf \hat n}_s=\kappa {\bf \hat t}$, 
and rescale the equation of motion via
\begin{align}
    {\bf r}&=L{\bf r}'~, \ \ \ z=Lz'~, \ \ \ s=Ls'~, \ \ \ \kappa = \frac{1}{L} \kappa' ~, \nonumber \\
    \Lambda &=\frac{A}{L^2} \Lambda' ~,  \ \ \ t= \frac{\zeta_\perp L^4}{A} t' ~, \ \ \ U= \frac{\zeta_\perp L^3}{A} U'.
\end{align}
If we let $\zeta_\perp/\zeta_\parallel \simeq 2$, and use 
${\bf \hat e}_x=\cos\theta{\bf \hat t} + \sin\theta{\bf \hat n}$ and
${\bf \hat e}_z=\sin\theta{\bf \hat t} - \cos\theta{\bf \hat n}$, and then drop the primes, we obtain
\begin{widetext}
\begin{equation}
    {\bf r}_t=\left(U\cos\theta +V\sin\theta\right){\bf \hat t}+\left(U\sin\theta-V\cos\theta\right) {\bf \hat n}
    +\left(\kappa_{ss}-\kappa^3+\kappa\Lambda\right){\bf \hat n}
    +\eta\left(3\kappa\kappa_s-\Lambda_s\right){\bf \hat t}~.
    \label{eom2}
\end{equation}
Differentiating \eqref{eom2} with respect to $s$ to obtain the equation of motion of
the tangent angle $\theta$ and that for $\Lambda$:
\begin{equation}
    \theta_t=-\theta_{ssss}+3\theta_s^2\theta_{ss}-\left(\theta_s\Lambda\right)_s  
     +2\left(3\theta_s^2\theta_{ss}-\theta_s\Lambda_s\right)-U_s\sin\theta+V_s\cos\theta~,
     \label{final1}
     \end{equation}
     while enforcing inextensibility leads to the equation for $\Lambda$,
     \begin{equation}
    \left(\partial_{ss}-\frac{1}{2}\theta_s^2\right)\Lambda=
    -\frac{1}{2}\theta_s^4+3\theta_{ss}^2+\frac{7}{2}\theta_s\theta_{sss}
    +\frac{M}{\eta}\sin(\theta)\cos(\theta)~.
    \label{final2}
\end{equation}
\end{widetext}
The boundary conditions at the attachment point are those of a clamped filament,
\begin{align}
    &\theta(0)=\frac{\pi}{2}~, \ \ \  \theta_{sss}(0)-\theta_s(0)^3+
    \theta_s(0)\Lambda(0)=0~, \\ \nonumber
    &\Lambda_s(0)-3\theta_s(0)\theta_{ss}(0)=0~.
\end{align}
whereas at the free end we have
\begin{equation}
    \theta_s(1)=0~, \ \ \ \theta_{ss}(1)=0~,\ \ \ \Lambda(1)=0~. \ \ \
\end{equation}

\subsection{Self-consistent model}

\begin{figure}[b]
    \includegraphics[width=0.98\columnwidth]{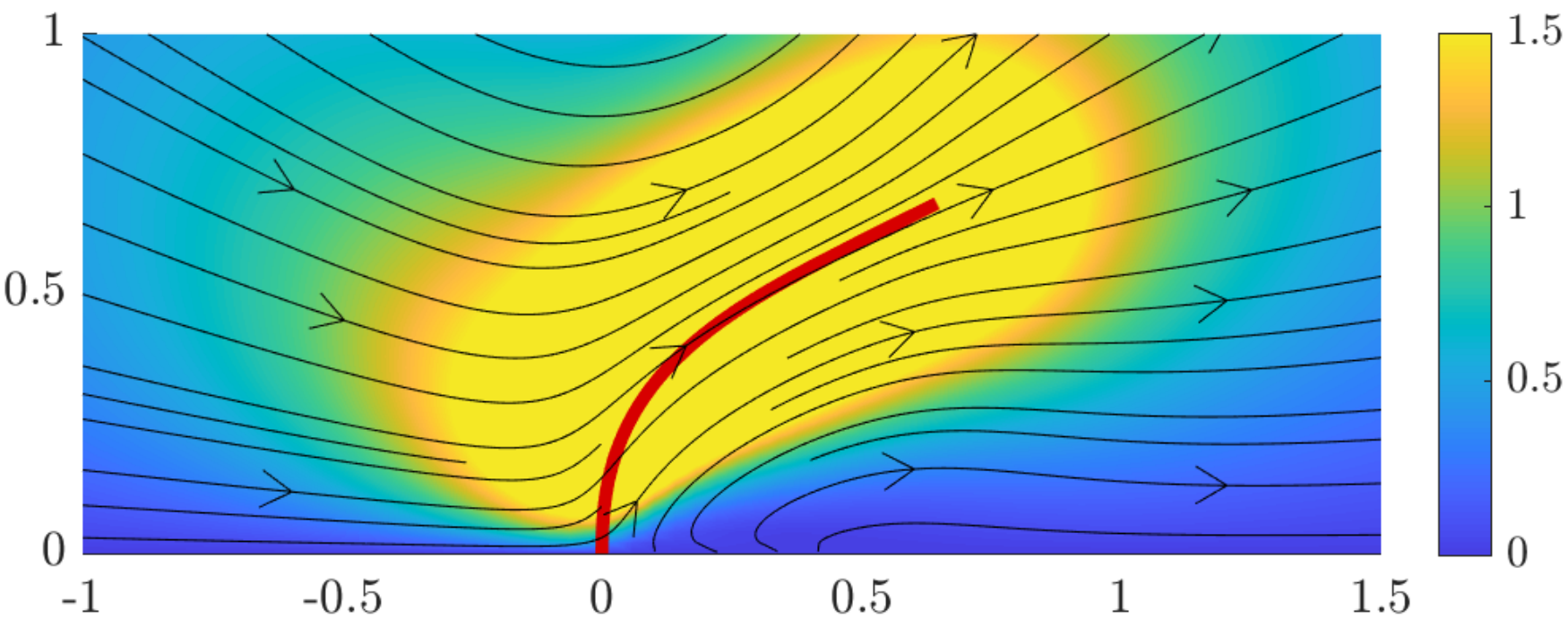}
    \caption{Flow field near a filament in the swirling regime.  A filament in the bent shape found in the
    spherical geometry of the main text was placed near a flat, no-slip wall and the resulting flow field 
    computed numerically.  Note the shear flow downstream.}
    \label{flow_near_filament}
\end{figure}

If we take the typical bent filament shape in the streaming regime (Fig. 2(d) of main text) and place
it near a flat no-slip wall, we can understand in the simplest situation the flow it produces.  As shown 
in the streamlines and velocity colormap of Figure \ref{flow_near_filament} there is a boundary-layer 
phenomenon near the wall that arises from the combination of tangential forcing of the flow by the
motors and the no-slip condition at the wall.  The most prominent part of the flow is a lobe of high speeds 
emanating from the bent portion of the filament, directed downstream.  This feature forms the basis
of the self-consistent model for the swirling transition.

As Fig. \ref{flow_near_filament} shows, the flow downstream from a bent filament is approximately
simple shear.  This can be seen directly in Blake's analysis 
\cite{Blake_SM} of the flow due to a point force near a no-slip wall.  For the purposes of a
simple self-consistent model, we consider only the asymptotic form of that flow evaluated at
${\bf x}=(x,y,z)$ due to a point force ${\bf F}$ at $(0,0,h)$:
\begin{align}
    u_i\simeq&\frac{F_k}{8\pi\mu}\biggl[\frac{12hx_ix_{\alpha}x_3\delta_{k\alpha}}{\vert{\bf x}\vert^5}\\ \nonumber
    & +h^2\delta_{k3}\left(-\frac{(12+6\delta_{i3})x_ix_3}{\vert{\bf x}\vert^5}+
    \frac{30x_ix_3^3}{\vert{\bf x}\vert^7}\right)\biggr]~.
\end{align}
Taking the leading-order term, and ${\bf F}$ along $x$, we obtain
\begin{equation}
    {\bf u}(x,z)=\dot\gamma(x) z {\bf \hat e}_x
\end{equation}
where the shear rate is
\begin{equation}
    \dot\gamma(x)=\frac{3Fh}{2\pi\mu x^3}~.
\end{equation}
In the notation used in the tangent angle dynamics \eqref{final1}, we have $U=Mz(s)$ and $V=0$, where 
\begin{equation}
    M=\frac{\eta\dot\gamma L^4}{A}\sim \frac{3\sigma}{c}\left(\frac{\rho}{\phi}\right)^{3/2}~.
\end{equation}
a control parameter that appears also in the dynamics elastic filaments in extensional flows 
\cite{YoungShelley_SM,Kantsler_SM}.  Here, the second relation uses the definitions of the dimensionless
control parameters $\sigma$ and $\rho$ from the main text.
To complete the self-consistency calculation, we set the $x$-component of the force ${\bf F}$ to be
proportional to the projection of the tangent vector at the distal tip, so 
$M \to M\sin(\theta(L))$

\begin{figure}[t]
	\includegraphics[width=0.98\columnwidth]{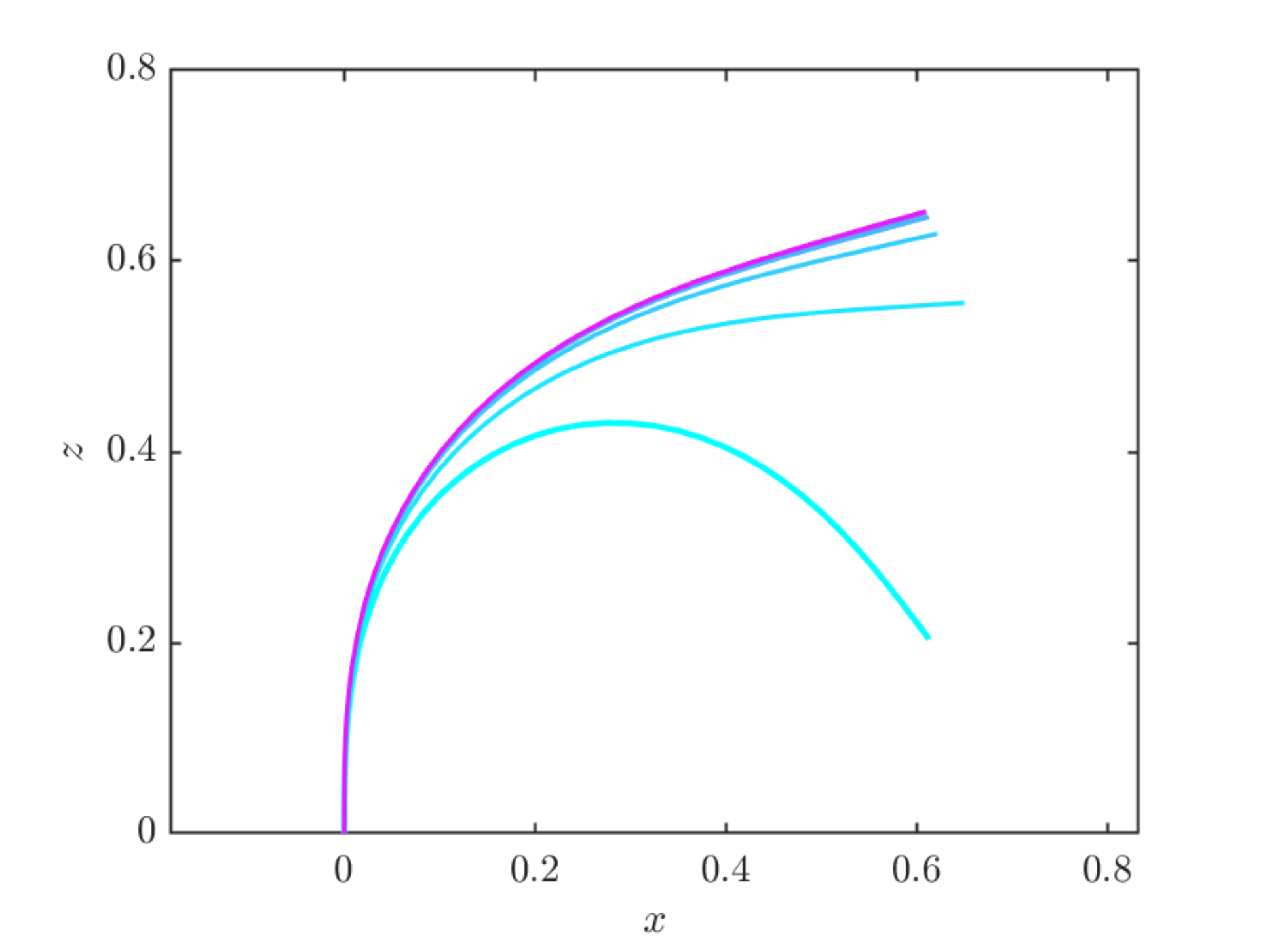}
	\caption{Results of the self-consistent calculation. Relaxation of a filament to steady bent configuration, 
	at $\sigma=45$ and $\rho=5$, with colors interpolating between highly bent initial condition (cyan) to
	final state (magenta) over a (dimensionless) time of $0.05$.}
	\label{filament_relaxation_sc}
\end{figure}

The tangent angle dynamics \eqref{final1} and \eqref{final2} 
were solved by a Crank-Nicholson method
via a pentadiagonal matrix for the fourth derivative, 
treating all the nonlinearities
explicitly, and discretizing the $\Lambda$ problem as a tridiagonal matrix in which
the diagonal term involving $\theta_s^2$ was updated after each time step.  The initial
condition was $\theta(0,s)=\pi/2-a\sin(\pi s/2)^2$, with $a=2.5$ and the time step
$dt=0.1 ds^4$, with $ds=1/(N-1)$, where $N$ is the total number of
grid points ($N=41$ is sufficient).

Figure \ref{filament_relaxation_sc} shows a typical result of the self-consistent 
calculation in the regime of stable streaming, illustrating how a highly bent initial condition relaxes
to a conformation like that seen in the calculations described in the main text (compare Figs. 3(c) from 
discrete filament calculations and
4(b) from continuum model). 

% A coarse parameter sweep results in the 
% numerical stability diagram shown in Fig. \ref{stability_diagram_sc}, which displays the same
% features as that from the continuum model, including a steep vertical asymptote at 
% $\sigma\sim 25$, a tongue of stable streaming at intermediate values, and collective oscillations
% at higher $\sigma$.  Given the simplifying assumptions made in the model \textemdash replacement of
% a fraction of the filament by a single stokeslet, approximation of the flow field as pure shear, etc. 
% \textemdash this level of agreement lends support to the heuristic picture of the swirling instability
% embodied in the self-consistent model.

%\begin{figure}[t]
%	\begin{center}
%	\includegraphics[width=0.98\columnwidth]{Swirling_figureS3.pdf}
%	\caption{Stability diagram (placeholder for final version).}
%	\end{center}
%	\label{stability_diagram_sc}
% \end{figure}

\section{Further Results from Continuum Model}

\subsection{Dynamics of the transition to swirling}

In Supplemental Video 1, we show the dynamics of the transition from the unstable equilibrium of radially oriented filaments to stable streaming flow, corresponding to Fig. 4b (main text). In this parameter regime, the filament array initially buckles, and large azimuthal flows are generated by rapid deformation of the array. As the filament motion slows, the filaments become oriented parallel to the cylinder walls. Azimuthal flows are continuously generated by the streaming mechanism, and the drag from these flows moving past the stationary filaments is sufficient to pin them into the observed conformation.

\subsection{Effect of cylindrical confinement}

\begin{figure}[b]
    \includegraphics[width=0.9\columnwidth]{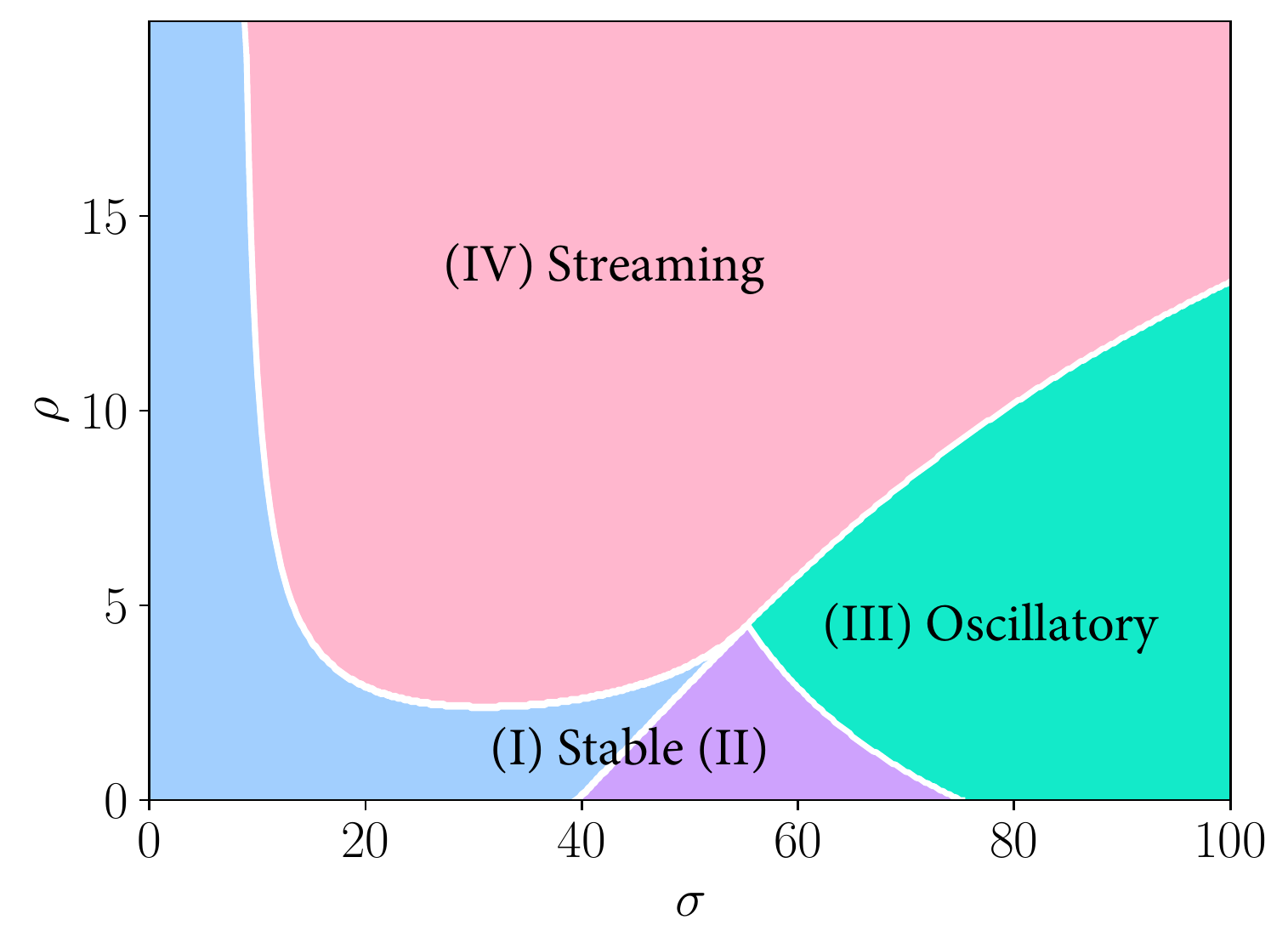}
    \caption{Results of the stability analysis in the planar geometry. Regions are colored and coded as in the main text.}
    \label{flat_phase}
\end{figure}

\begin{figure*}[t]
    \includegraphics[width=1.8\columnwidth]{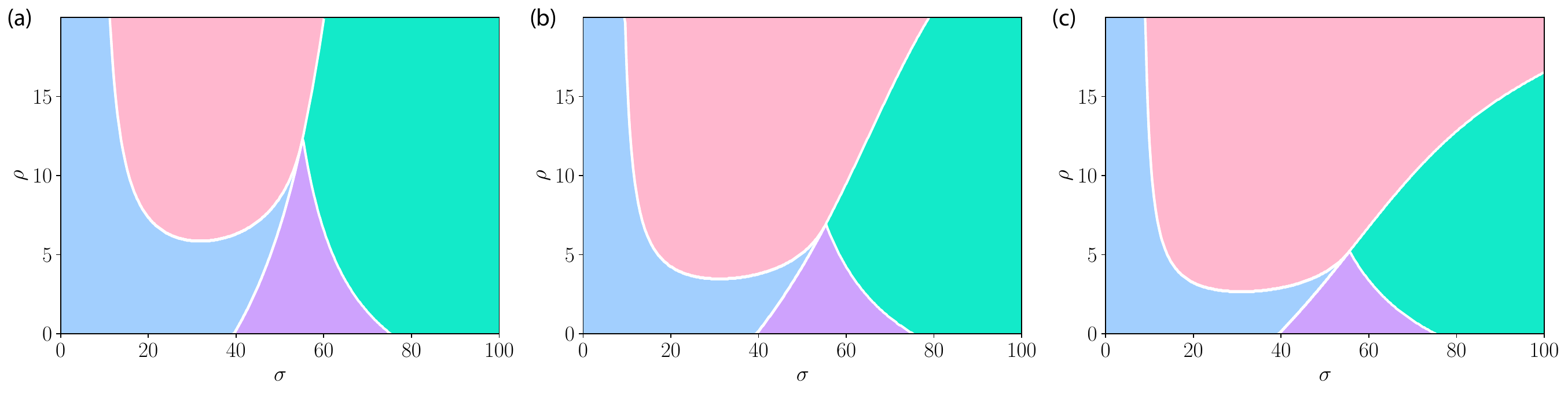}
    \caption{Results of the stability analysis in the confined geometry. The confinement ratios are 2.2:1 (Panel a), 4:1 (Panel b), and 20:1 (Panel c). See Fig.~\ref{flat_phase} for a description of the different regions.}
    \label{cylindrical_phase}
\end{figure*}

In the main text, we present the evolution equation for a small perturbation to the equilibrium solution in a planar geometry. This equation:
\begin{equation}
   g_t = -g_{zzzz} -\sigma\left[(1-z)g_{z}\right]_z + \rho
   \left[\sigma(1-z)g+g_{zz}\right],
    \label{linstabSM}
\end{equation}
remains local due to the simplicity of the Stokes flow, allowing the individual terms to be easily identified and interpreted. In order to better capture the real geometry of the oocyte, in the main text we present the results of the linear stability analysis in a cylindrical geometry (Fig 4a, main text). The same analysis, over the same range of effective densities ($\rho$) and effective motor force-densities ($\sigma$) is shown in Fig. \ref{flat_phase}, computed for the planar geometry. These results are qualitatively the same as those shown in the main text in Fig 4a, indicating that the essential structure of the swirling transition is due to the interactions between nearby filaments in the fiber bed, and does not depend in a critical way on the specific geometry of the oocyte.

In, Fig. \ref{cylindrical_phase} we present the same analysis for different confinement ratios --- 2.2:1 (Fig \ref{cylindrical_phase}a), 4:1 (Fig \ref{cylindrical_phase}b), and 20:1, (Fig \ref{cylindrical_phase}c). Note that the ratio 2.2:1 indicates a cylinder diameter of $2.2L$, which is extremely confined, with filaments reaching $\approx91\%$ of the way to the center of the cylinder. Even under such strong confinement, the qualitative structure remains, although the quantitative value at which the bifurcations occur changes significantly. The effect of strong confinement is to suppress the appearance of streaming at larger forcing values. This makes sense: when forcing is large, the tendency of lone fibers is to oscillate. In the planar system, when fibers are sufficiently dense, the strong flows caused by motors walking on neighboring fibers is sufficient to pin the fiber into a stable deformed state, suppressing oscillations (in fact, the drag from these flows produces extensile forces on the fibers, reducing the net contractile forcing generated by the motors). Under high confinement, the strength of such flows near the fiber tips is reduced, and the flow can no longer suppress the tendency of the fiber to oscillate. When the confinement is not so extreme, even the quantitative values change little, from 10:1 (Fig 4a, main text), to 20:1 (Fig \ref{cylindrical_phase}c), to the flat geometry (Fig \ref{flat_phase}).

It may surprise the reader that when $\rho\to0$, the behavior (as a function of $\sigma$) at all confinement ratios converges to \emph{exactly} the same thing. This is because the continuum model imposes boundary conditions on the coarse-grained flow only; as $\rho\to0$ this flow goes to $0$ with it and the dynamics are those of a fiber moving in a quiescent background.

\begin{figure}[b]
    \includegraphics[width=0.9\columnwidth]{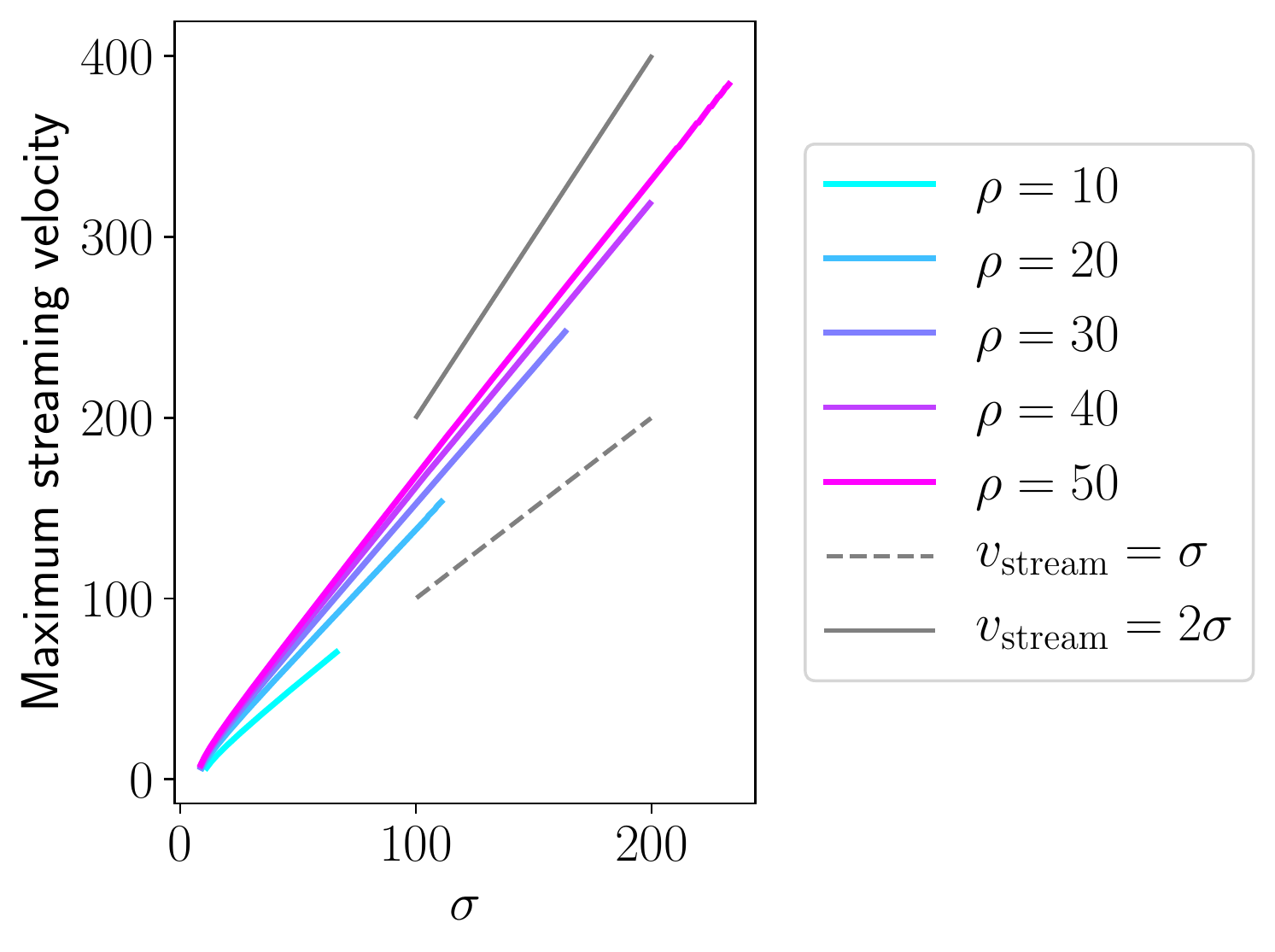}
    \caption{Dependence of streaming speed on the force density ($\sigma$) generated by kinesin-1 motors. When $\rho$ is large, the streaming speed approaches the estimate $v_\textnormal{streaming}=2\sigma$.}
    \label{streaming_speeds}
\end{figure}

\subsection{Streaming Speeds in Dense Arrays}

By solving for $-\r_{ssss} + (\Lambda\r_s)_s$ in Eq. 2 (main text), substituting into Eq. 3 (main text), defining $\xi\equiv\chi_\textnormal{MT}\rho\mathcal{J}^{-1}$, and ignoring frame transformations for the sake of simplicity, we find that:
\begin{equation}
    -\grad^2\u + \grad p + \xi(\mathbb{I}-\r_s\r_s/2)(\u-\r_t) = \xi\sigma\r_s.
    \label{eq:simple_balance}
\end{equation}
This is a forced Brinkman equation with an anisotropic permeability. When $\xi$ is large, the skeletal drag term dominates the left-hand side, and at steady state $\r_t=0$. We may thus approximate the steady streaming velocity as simply $\u \approx \sigma(\mathbb{I}+\r_s\r_s)\r_s = 2\sigma\r_s$. For homogeneous flows in the planar geometry, and axisymmetric flows in the cylindrical geometry, the flow is purely in the $\hat x$ or $\hat\theta$ (azimuthal) direction, respectively. Since $\r_s$ is not purely coincident with $\hat x$ or $\hat\theta$, some of the forcing generated by the molecular motors must be absorbed into the pressure gradient, and so we take this estimate to be an upper bound: $v_\textnormal{streaming}\leq2\sigma$. The maximum steady-state streaming speed as a function of $\sigma$, as computed by the continuum model in 10:1 confinement, is shown in Fig.~\ref{streaming_speeds}, for $\rho=10,\,20,\,\ldots,\,50$. For smaller values of $\rho$, the fiber is less deformed and drag contributes less to the balance in Eq.~\ref{eq:simple_balance}, leading to a relationship closer to $v_\textnormal{streaming}=\sigma$. When $\rho$ is larger, drag dominates Eq.~\ref{eq:simple_balance}, the fiber is nearly azimuthally aligned, and the streaming speed approaches the bound $v_\textnormal{streaming}=2\sigma$.

In dimensional units, this bound is $v_\textnormal{streaming}\leq 2\sigma A/\eta L^3$. In the main text, we used a simple argument based on Darcy's law to estimate $v_\textnormal{streaming}=(\sigma A/\eta L^3)(8\pi/c)$. The leading factor of two in the more refined estimate presented in this SI comes from taking into account fiber anisotropy and the nearly azimuthal alignment of the fibers, while the disappearance of the factor of $8\pi/c$ arises from properly accounting for the fibers aspect ratio. In either case, the disappearance of the MT density from the estimate of streaming velocity arises due to the fact that both the forcing ($\xi\sigma\r_s$) and the skeletal drag ($\xi(\mathbb{I}-\r_s\r_s/2))$ scale with the density $\rho$.


%merlin.mbs apsrev4-1.bst 2010-07-25 4.21a (PWD, AO, DPC) hacked
%Control: key (0)
%Control: author (8) initials jnrlst
%Control: editor formatted (1) identically to author
%Control: production of article title (-1) disabled
%Control: page (0) single
%Control: year (1) truncated
%Control: production of eprint (0) enabled
\begin{thebibliography}{0}%
\makeatletter
\providecommand \@ifxundefined [1]{%
 \@ifx{#1\undefined}
}%
\providecommand \@ifnum [1]{%
 \ifnum #1\expandafter \@firstoftwo
 \else \expandafter \@secondoftwo
 \fi
}%
\providecommand \@ifx [1]{%
 \ifx #1\expandafter \@firstoftwo
 \else \expandafter \@secondoftwo
 \fi
}%
\providecommand \natexlab [1]{#1}%
\providecommand \enquote  [1]{``#1''}%
\providecommand \bibnamefont  [1]{#1}%
\providecommand \bibfnamefont [1]{#1}%
\providecommand \citenamefont [1]{#1}%
\providecommand \href@noop [0]{\@secondoftwo}%
\providecommand \href [0]{\begingroup \@sanitize@url \@href}%
\providecommand \@href[1]{\@@startlink{#1}\@@href}%
\providecommand \@@href[1]{\endgroup#1\@@endlink}%
\providecommand \@sanitize@url [0]{\catcode `\\12\catcode `\$12\catcode
  `\&12\catcode `\#12\catcode `\^12\catcode `\_12\catcode `\%12\relax}%
\providecommand \@@startlink[1]{}%
\providecommand \@@endlink[0]{}%
\providecommand \url  [0]{\begingroup\@sanitize@url \@url }%
\providecommand \@url [1]{\endgroup\@href {#1}{\urlprefix }}%
\providecommand \urlprefix  [0]{URL }%
\providecommand \Eprint [0]{\href }%
\providecommand \doibase [0]{http://dx.doi.org/}%
\providecommand \selectlanguage [0]{\@gobble}%
\providecommand \bibinfo  [0]{\@secondoftwo}%
\providecommand \bibfield  [0]{\@secondoftwo}%
\providecommand \translation [1]{[#1]}%
\providecommand \BibitemOpen [0]{}%
\providecommand \bibitemStop [0]{}%
\providecommand \bibitemNoStop [0]{.\EOS\space}%
\providecommand \EOS [0]{\spacefactor3000\relax}%
\providecommand \BibitemShut  [1]{\csname bibitem#1\endcsname}%
\let\auto@bib@innerbib\@empty
%</preamble>
\end{thebibliography}%


\begin{thebibliography}{99}

\bibitem{NeedlemanShelley}
D. Needleman and M.J. Shelley, The stormy fluid dynamics of the living cell,
\href{https://doi.org/10.1063/PT.3.4292}{Physics Today {\bf 72}, 32 (2019)}.

\bibitem{Drosophila_review1} A short primer on the subject is:
R. Bastock and D. St. Johnston, \textit{Drosophila} oogenesis,
\href{https://doi.org/10.1016/j.cub.2008.09.011}{Curr. Biol. {\bf 18}, R1082 (2008)}.

\bibitem{HeWangMontell} L. He, X. Wang, and D.J. Montell, Shining light on {\it Drosophila} oogenesis:
live imaging of egg development, 
\href{https://doi.org/10.1016/j.gde.2011.08.011}{Curr. Op. Gen. \& Dev. {\bf 21}, 612 (2011)}.

\bibitem{streaming} The phenomenon of streaming was first discovered in plants, as reported by
B. Corti, {\it Osservazione Microscopische sulla Tremella e sulla Circulazione del Fluido
in Una Planto Acquaguola} (Appresso Giuseppe Rocchi, Lucca, Italy, 1774).

\bibitem{vandeMeent_review}
R.E. Goldstein and J.-W. van de Meent,
A Physical Perspective on Cytoplasmic Streaming,
\href{https://doi.org/10.1098/rsfs.2015.0030}{Interface Focus {\bf 5}, 20150030 (2015)}.

\bibitem{Gutzeit} H. Gutzeit and R. Koppa, Time-lapse film analysis of cytoplasmic streaming during
late oogenesis of {\it Drosophila}, 
\href{https://doi.org/}{J. Embryol. Exp. Morphol. {\bf 67}, 101 (1982)}.

\bibitem{Theurkauf92} W. Theurkauf, S. Smiley, M. Wong, and B. Alberts, Reorganization of the cytoskeleton
during {\it Drosophila} oogenesis: implications for axis specification and intercellular
transport
\href{https://doi/org/}{Development {\bf 115}, 923 (1992)}.

\bibitem{Theurkauf94} W.E. Theurkauf, Premature microtubule-dependent cytoplasmic streaming in
cappuccino and spire mutant oocytes, 
\href{https://doi/org/10.1126/science.8091233}{Science {\bf 265} 2093 (1994)}.

\bibitem{Palacios02} I.M. Palacios and D. St. Johnston, {\it Kinesin light chain}-independent 
function of the {\it Kinesin Heavy Chain} in cytoplasmic streaming and posterior localisation in 
the {\it Drosophila} oocyte,
\href{https://doi/org/10.1242/dev.00119}{Development 129:5473–5485 (2002)}.

\bibitem{Serbus05} L.R. Serbus, B.J. Cha, W.E. Theurkauf, W.M. Saxton, Dynein and the actin 
cytoskeleton control kinesin-driven cytoplasmic streaming in {\it Drosophila} oocytes,
\href{https://doi/org/10.1242/dev.01956}{ Development {\bf 132}, 3743 (2005)}.

\bibitem{Ganguly} S. Ganguly, L.S. Williams, I.M. Palacios, and R.E. Goldstein,
Cytoplasmic Streaming in \textit{Drosophila} Oocytes Varies with Kinesin Activity
and Correlates With the Microtubule Cytoskeleton Architecture, 
\href{https://doi.org/10.1073/pnas.1203575109}{Proc. Natl. Acad. Sci. USA {\bf 109}, 15109 (2012)}.

\bibitem{Drechsler1} M. Drechsler, F. Giavazzi, R. Cerbino, and I.M. Palacios, 
Active diffusion and advection in \textit{Drosophila} oocytes results from the interplay
of actin and microtubules, 
\href{https://doi.org/10.1038/s41467-017-01414-6}{Nat. Comm. {\bf 8}, 1520 (2017)}.

\bibitem{Drechsler2} M. Drechsler, L.F. Lang, L. Al-Khatib, H. Dirks, M. Burger, C.-B. Sch{\"o}nlieb and
I.M. Palacios, Optical flow analysis reveals that Kinesin-mediated advection impacts the orientation of
microtubules in the {\it Drosophila} oocyte, 
\href{https://doi.org/10.1091/mbc.E19-08-0440}{Mol. Biol. Cell {\bf 31}, 1246 (2020)}.

\bibitem{Monteith} C.E. Monteith, M.E. Brunner, I. Djagaeva, A.M. Bielecki, J.M. Deutsch,
and W.M. Saxton, A Mechanism for Cytoplasmic Streaming: Kinesin-Driven Alignment of 
Microtubules and Fast Fluid Flows, 
\href{https://doi.org/10.1016/j.bpj.2016.03.036}{Biophys. J. {\bf 110}, 2053 (2016)}.

\bibitem{KhucTrong_PRL}
P. Khuc Trong, J. Guck, and R.E. Goldstein,
Coupling of Active Motion and Advection Shapes Intracellular Cargo Transport,
\href{https://doi.org/10.1103/PhysRevLett.109.028104}{Phys. Rev. Lett. {\bf 109}, 028104 (2012)}.

\bibitem{Williams} L.S. Williams, S. Ganguly, P. Loiseau, B.F. Ng and I.M. Palacios, 
The auto-inhibitory domain and the ATP-independent microtubule-binding
region of Kinesin Heavy Chain are major functional domains for
transport in the \textit{Drosophila} germline, 
\href{https://doi.org/10.1242/dev.097592}{Development {\bf 141}, 176 (2014)}.

\bibitem{KhucTrong_eLife}
P. Khuc Trong, H. Doerflinger, J. Dunkel, D. St. Johnston, and R.E. Goldstein,
Cortical Microtubule Nucleation Can Organise the Cytoskeleton of
\textit{Drosophila} Oocytes to Define the Anteroposterior Axis, 
\href{https://doi.org/10.7554/eLife.06088}{eLife {\bf 4}, e06088 (2015)}.

\bibitem{Gelfand_jcb} W. Lu, M. Lakonishok, A.S. Serpenskaya, D. Kirchenb{\"u}echler, 
S.-C. Ling 
and V.I. Gelfand, Ooplasmic flow cooperates with transport and anchorage in {\it Drosophila} oocyte
posterior determination,
\href{https://doi.org/10.1083/jcb.201709174}{J. Cell. Biol. {\bf 217}, 3497 (2018)}.

\bibitem{Gittes} F. Gittes, B. Mickey, J. Nettleton and J. Howard, Flexural Rigidity of 
Microtubules and Actin Filaments Measured from Thermal Fluctuations in Shape,
\href{https://doi.org/10.1083/jcb.120.4.923}{J. Cell Biol. {\bf 120}, 923 (1993)}.

\bibitem{motorforce} K. Visscher, M.J. Schnitzer, and S.M. Block, Single Kinesin Molecules
Studied With a Molecular Force Clamp,
\href{https://doi.org/10.1038/22146}{Nature {\bf 400}, 184 (1999)}.

\bibitem{followerforce} G. Herrmann and R.W. Bungay RW, On the stability of
elastic systems subjected to nonconservative forces,
\href{https://doi.org/10.1115/1.3629660}{J. Appl. Mech. {\bf 31}, 435 (1964)}. 

\bibitem{BaylyDutcher} P.V. Bayly and S.K. Dutcher, Steady dynein forces induce flutter 
instability and propagating waves in mathematical models of flagella, 
\href{https://doi.org/10.1098/rsif.2016.0523}{J. R. Soc. Interface {\bf 13}, 20160523 (2016)}.

\bibitem{deCanio} G. De Canio, E. Lauga, and R.E. Goldstein, Spontaneous oscillations of elastic
filaments induced by molecular motors, 
\href{https://doi.org/10.1098/rsif.2017.0491}{J. R. Soc. Interface {\bf 14}, 20170491 (2017)}.

\bibitem{kanso_follower} F. Ling, H. Guo, and E. Kanso, Instability-driven oscillations of elastic microfilaments,
\href{https://doi.org/10.1098/rsif.2018.0594}{J. R. Soc. Interface {\bf 15}, 149 (2018).}

\bibitem{YoungShelley} Y.-N. Young and M.J. Shelley, Stretch-Coil Transition and 
Transport of Fibers in Cellular Flows,
\href{https://doi.org/10.1103/PhysRevLett.99.058303}{Phys. Rev. Lett. {\bf 99}, 058303 
(2007)}.

\bibitem{Kantsler} V. Kantsler and R.E. Goldstein, Flucutations, Dynamics, and the
Stretch-Coil Transition of Single Actin Filaments in Extensional Flows,
\href{https://doi.org/10.1103/PhysRevLett.108.038103}{Phys. Rev. Lett. 
{\bf 108}, 038103 (2012)}.

\bibitem{SteinShelley}
D.B. Stein and M.J. Shelley,
Coarse graining the dynamics of immersed and driven fiber assemblies,
\href{https://doi.org/10.1103/PhysRevFluids.4.073302}{Phys. Rev. Fluids {\bf 4}, 073302 (2019)}.

\bibitem{flowfield} I.M. Palacios and M. Drechsler, private communication (2020), based on methods detailed
earlier \cite{Ganguly,Williams}.

\bibitem{Gelfand} W. Lu, M. Winding, M. Lakonishok, J. Wildonger and V.I. Gelfand, Microtuble-microtubule 
sliding by kinesin-1 is essential for normal cytoplasmic streaming in {\it Drosophila} oocytes,
\href{https://doi.org/10.1073/pnas.1522424113}{Proc. Natl. Acad. Sci. USA {\bf 113}, E4995 (2016)}.

\bibitem{KellerRubinow} J.B. Keller and S.I. Rubinow, Slender-body theory for slow viscous flow,
\href{https://doi.org/10.1017/S0022112076000475}{J. Fluid Mech. {\bf 75}, 705 (1976)}.

\bibitem{RFT} J. Gray and G.J. Hancock, The propulsion of sea-urchin spermatozoa, 
J. Exp. Biol.{\bf 32}, 802 (1955).

\bibitem{TornbergShelley} A.K. Tornberg, and M.J. Shelley, Simulating the dynamics and interactions of flexible fibers in Stokes flows, \href{https://doi.org/10.1016/j.jcp.2003.10.017}{J. Comput. Phys. {\bf 196}, 1 (2004)}.

\bibitem{GL} R.E. Goldstein and S.A. Langer, Nonlinear dynamics of stiff polymers,
\href{https://doi.org/10.1103/PhysRevLett.75.1094}{Phys. Rev. Lett. {\bf 75}, 1094 (1995)}.

\bibitem{SM}  See Supplemental Material at \url{http://link.aps.org/supplemental/xxx} for
further details and results.

\bibitem{thesis} Further examples and details are in: 
G. De Canio, Motion of filaments induced by molecular motors:
from individual to collective dynamics, PhD thesis, University of Cambridge (2018).

\bibitem{MaulKim} C. Maul and S. Kim, Image systems for a Stokeslet inside a rigid 
spherical container,
\href{https://doi.org/10.1063/1.868223}{Phys. Fluids {\bf 6}, 2221 (1994)}.

\bibitem{synchro} R.E. Goldstein, Green Algae as Model Organisms for Biological Fluid Dynamics,
\href{https://doi.org/10.1146/annurev-fluid-010313-141426}{Annu. Rev. Fluid Mech. {\bf 47}, 343 (2015)}.

\bibitem{NEL} T. Niedermayer, B. Eckhardt and P. Lenz, Synchronization, phase locking, and metachronal
wave formation in ciliary chains, 
\href{https://doi.org/10.1063/1.2956984}{Chaos {\bf 18}, 037128 (2008)}.

\bibitem{Yotsuyanagi} Y. Yotsuyanagi, Recherches sur les ph{\'e}énomen{\`e}s moteurs dans les fragments de 
protoplasme isol{\'e}s. I. Mouvement rotatoire et le processus de son apparition,
\href{https://doi.org/10.1508/cytologia.18.146}{Cytologia {\bf 18}, 146 (1953)};
Recherches sur les ph{\'e}énomen{\`e}s moteurs dans les fragments de 
protoplasme isol{\'e}s. II. Mouvements divers d{'e}termin{'e}s par la condition de milieu,
\href{https://doi.org/10.1508/cytologia.18.202}{Cytologia {\bf 18}, 202 (1953)}.

\bibitem{confinement_expt} H. Wioland, F.G. Woodhouse, J. Dunkel, J.O. Kessler, and R.E. Goldstein,
Confinement Stabilizes a Bacterial Suspensions into a Spiral Vortex,
\href{https://doi.org/10.1103/PhysRevLett.110.268102}{Phys. Rev. Lett. {\bf 110}, 268102 (2013)}.

\bibitem{Saintillan} D. Saintillan and M.J. Shelley, Instabilities and Pattern Formation in Active Particle Suspensions:
Kinetic Theory and Continuum Simulations, 
\href{https://doi.org/10.1103/PhysRevLett.100.178103}{Phys. Rev. Lett. 100, 178103 (2008).}

\bibitem{Saintillan2018extensile} D. Saintillan, M. J. Shelley, A. Zidovska, Extensile motor activity drives coherent motions in a model of interphase chromatin, \href{https://doi.org/10.1073/pnas.1807073115}{Proc. Natl. Acad. Sci. USA {\bf 115}, 11442 (2018).}

\bibitem{confinement_theory} F.G. Woodhouse and R.E. Goldstein, Spontaneous Circulation of 
Confined Active Suspensions, 
\href{https://doi.org/10.1103/PhysRevLett.109.168105}{Phys. Rev. Lett. {\bf 109}, 168105 (2012)}.

\bibitem{NRZS} E. Nazockdast, A. Rahimian, D. Zorin, and M.J. Shelley, A fast platform for simulating semi-flexible fiber suspensions applied to cell mechanics, \href{https://doi.org/10.1016/j.jcp.2016.10.026}{J. Comput. Phys. {\bf 329} (2017)}.

\bibitem{methods} Differentiation of \eqref{eq:continuum:fiber} yields an equivalent equation 
for the tangent-vector field $\hatt$ \cite{SteinShelley}, which we have found to be 
numerically more stable.

\bibitem{Landau} L.D. Landau and E.M. Lifshitz, \textit{Theory of Elasticity}, 
2nd ed. (Pergamon Press, Oxford, 1970), p. 99, Problem 7.

\bibitem{ponytails} R.E. Goldstein, P.B. Warren and R.C. Ball, Shape of a Ponytail and the
Statistical Physics of Hair Fiber Bundles,
\href{https://doi.org/10.1103/PhysRevLett.108.078101}{Phys. Rev. Lett. {\bf 108}, 078101 (2012)}.

\bibitem{Loiseau} P. Loiseau,, R. Davies, L.S. Williams, M. Mishima and I.M. Palacios, 
\textit{Drosophila} PAT1 is required for Kinesin-1 to transport cargo and to maximize its
motility, \href{https://doi.org/10.1242/dev.048108}{Development {\bf 137}, 2763 (2010)}.

\bibitem{eLife} D.R. Brumley, K.Y. Wan, M. Polin and R.E. Goldstein, Flagellar synchronization through 
direct hydrodynamic interactions,
\href{https://doi.org/10.7554/eLife.02750}{eLife {\bf 3}, e02750 (2014)}.

\bibitem{Blake} J.R. Blake, A note on the image system for a stokeslet in a no-slip boundary, 
\href{https://doi.org/10.1017/S0305004100049902}{Math. Proc. Camb. Phil. Soc. {\bf 70}, 303 (1971)}.

\bibitem{lauga_pillars} J.-B. Thomazo, E. Lauga, B. Le R{\'e}v{\'e}rend, E. Wandersman,
and A.M. Prevost, Collective stiffening of soft hair assembles, 
\href{https://www.arxiv.org/abs/2002.02834}{arXiv:2002.02834}.

\end{thebibliography}

\begin{thebibliography}{99}

\bibitem{MK94_SM} C. Maul and S. Kim, Image systems for a Stokeslet inside a rigid 
spherical container,
\href{https://doi.org/10.1063/1.868223}{Phys. Fluids {\bf 6}, 2221 (1994)}.

\bibitem{Blake_SM} J.R. Blake, A note on the image system for a stokeslet in a no-slip boundary, 
\href{https://doi.org/10.1017/S0305004100049902}{Math. Proc. Camb. Phil. Soc. {\bf 70}, 303 (1971)}.

\bibitem{YoungShelley_SM} Y.-N. Young and M.J. Shelley, Stretch-Coil Transition and 
Transport of Fibers in Cellular Flows,
\href{https://doi.org/10.1103/PhysRevLett.99.058303}{Phys. Rev. Lett. {\bf 99}, 058303 
(2007)}.

\bibitem{Kantsler_SM} V. Kantsler and R.E. Goldstein, Flucutations, Dynamics, and the
Stretch-Coil Transition of Single Actin Filaments in Extensional Flows,
\href{https://doi.org/10.1103/PhysRevLett.108.038103}{Phys. Rev. Lett. 
{\bf 108}, 038103 (2012)}.

\end{thebibliography}
\end{document}